\documentclass[runningheads]{svmult}

\usepackage{makeidx}   
\usepackage{graphicx}  
\usepackage{subeqnar}  
\usepackage{multicol}  
\usepackage{physprbb}  
\makeindex             

\usepackage{amsmath}   
\usepackage{amssymb}
\usepackage{xspace}
\newcommand{\powersep}{{\ensuremath{\times}}}
\newcommand{\Msun}{{\ensuremath{\mathrm{M}_{\odot}}}\xspace}

\newcommand{\Ep}[1]{{\ensuremath{10^{#1}}}}
\newcommand{\EE}[1]{{\ensuremath{\powersep\Ep{#1}}}}
\newcommand{\isofont}[1]{{\mathrm{#1}}}
\newcommand{\isomass}[1]{{\ensuremath{\isofont{^{#1}}}}}
\newcommand{\isocharge}[1]{{\ensuremath{\isofont{_{#1}}}}}
\newcommand{\isotope}[3]{{\ensuremath{\isocharge{#1}\isomass{#2}\isofont{#3}}}}

\newcommand{\II}[2]{{\isotope{}{#1}{#2}}}
\newcommand{\Ob}{{\ensuremath{\Omega_{\mathrm{b}}}}\xspace}
\newcommand{\Om}{{\ensuremath{\Omega_{\mathrm{m}}}}\xspace}
\newcommand{\Ol}{{\ensuremath{\Omega_{\Lambda}}}\xspace}
\newcommand{\Ho}{{\ensuremath{H_0}}\xspace}
\newcommand{\fst}{{\ensuremath{f_{\mathrm{1st}}}}\xspace}
\newcommand{\Mst}{{\ensuremath{M_{\mathrm{1st}}}}\xspace}
\newcommand{\rpsn}{{\ensuremath{r_{\mathrm{2SN}}}}\xspace}
\newcommand{\rhoc}{{\ensuremath{\rho_{\mathrm{c}}}}\xspace}
\newcommand{\dL}{{\ensuremath{d_{\mathrm{lum}}}}\xspace}
\newcommand{\dLs}{{\ensuremath{d_{\mathrm{lum}}^{\,2}}}\xspace}
\begin{document}
\title*{Evolution and Explosion of Very Massive Primordial Stars}
\toctitle{Very Massive Primordial Stars}
\titlerunning{Very Massive Primordial Stars}
%
\author{Alexander Heger\inst{1}
\and S. E.\ Woosley\inst{1}
\and Isabelle Baraffe\inst{2}
\and Tom Abel\inst{3}}
\authorrunning{Alexander Heger et al.}
%
%
\institute{
Department of Astronomy and Astrophysics, 
University of California,
1156 High Street,
Santa Cruz, CA 95064, U.S.A.  
\and 
Ecole Normale Sup\'erieure, C.R.A.L (UMR 5574 CNRS), 69364 Lyon
Cedex 07, France
\and
Institute of Astronomy, Madingley Road, Cambridge, CB3 0HA, England 
}

\maketitle              

\begin{abstract}
While the modern stellar IMF shows a rapid decline with increasing
mass, theoretical investigations suggest that very massive stars
($\gtrsim100\,\Msun$) may have been abundant in the early universe.
Other calculations also indicate that, lacking metals, these same
stars reach their late evolutionary stages without appreciable mass
loss.  After central helium burning, they encounter the
electron-positron pair instability, collapse, and burn oxygen and
silicon explosively.  If sufficient energy is released by the burning,
these stars explode as brilliant supernovae with energies up to 100
times that of an ordinary core collapse supernova. They also eject up
to 50\,\Msun of radioactive $^{56}$Ni.  Stars less massive than
140\,\Msun or more massive than 260\,\Msun should collapse into black
holes instead of exploding, thus bounding the pair-creation supernovae
with regions of stellar mass that are nucleosynthetically sterile.
Pair-instability supernovae might be detectable in the near infrared
out to redshifts of 20 or more and their ashes should leave a
distinctive nucleosynthetic pattern.
\end{abstract}

\section{Introduction}

Owing to the lack of any metals, the cooling processes that govern
star formation are greatly reduced for first generation of stars (Pop
III).  Magnetic fields and turbulence may also be less important at
these early times \cite{ABN00}.  Consequently, theoretical studies
\cite{Lar98} indicate that the Jeans mass for primordial stars in
their special environment may have been as great as $\sim1000\,\Msun$.
Numerical simulation of primordial star formation predict the
occurrence of such stars at red shifts $\sim20$ and an initial mass
function (IMF) that either peaks at $\sim100\,\Msun$
\cite{ABN00,BCL99} or is bimodal \cite{NU00}, i.e., also contains
stars of a few \Msun.

\begin{figure}
\includegraphics[angle=270,width=\columnwidth]{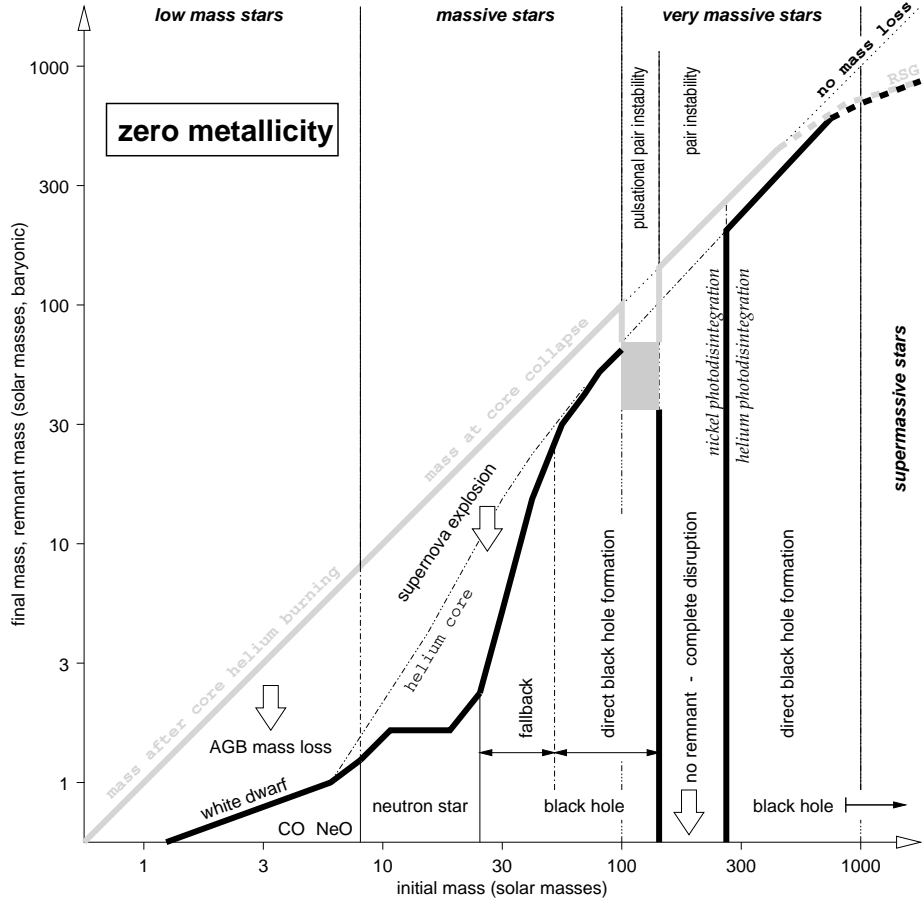}
\caption{Initial-final mass function of non-rotating Pop III stars.
The \textsl{x-axis} gives the initial mass.  The
\textsl{y-axis} gives both the final mass of the collapsed remnant
(\textsl{thick black curve}) and the mass of the star when the event
begins that produces that remnant (e.g., mass loss in AGB stars,
supernova explosion for those stars that make a neutron star, etc.;
\textsl{thick gray curve}).  We distinguish four regimes of initial
mass: \emph{low mass stars} below $\sim10\,\Msun$ that end as white
dwarfs; \emph{massive stars} between $\sim10\,\Msun$ and
$\sim100\,\Msun$ that form an iron core that eventually collapses;
\emph{very massive stars} between $\sim100\,\Msun$ and
$\sim1000\,\Msun$ that encounter the pair instability; and
\emph{supermassive stars} (arbitrarily) above $\sim1000\,\Msun$.
Since no mass loss is expected for $Z=0$ stars before the final stage,
the grey curve is approximately the same as the line of no mass loss
(\textsl{dotted}).  Exceptions are $\sim100-140\,\Msun$ where the
pulsational pair instability ejects the outer layers of the star
before it collapses, and above $\sim500\,\Msun$ where pulsational
instabilities in red supergiants may lead to significant mass loss
\cite{BHW01}.  Since the magnitude of the latter is uncertain, lines
are drawn \textsl{dashed}.  For a more detailed description, please
refer to
\cite{HW01}\label{MM3}}
\end{figure}

Once formed, at solar metallicity, massive stars ordinarily experience
significant mass loss \cite{Fig98} and may end as relatively small
objects, but for low metallicity mass loss is suppressed.  In
\S~\ref{loss} we discuss the peculiarities of mass loss and evolution
of very massive primordial stars.  Figure~\ref{MM3} gives an overview
of expected final fates of metal-free stars as a function of initial
mass.  In \S~\ref{z20} we examine the expected light curve of a
pair-creation supernova from a 250\,\Msun star at a redshift of $z=20$
and in \S~\ref{nuc} we review nucleosynthetic yields from Pop III.
Some conclusions are given in \S~\ref{end}.

\section{Mass Loss}
\label{loss}

Current studies of winds driven from hot stars by interactions with
atomic lines indicate that their mass loss rate decreases with
metallicity as $Z^{1/2}$ \cite{Kud00,KP00} or even more \cite{VKL01}.
This scaling seems to hold down to 0.1\,\% solar metallicity.
Extrapolating to zero, it seems reasonable that mass loss from this
wind driving mechanism becomes negligible.  Winds driven by continuum
opacity in low metal stars are not very well understood (Kudritzki,
priv.\ com.), and will be neglected here.

\begin{figure}[t]
\centering
\includegraphics[width=0.67\columnwidth]{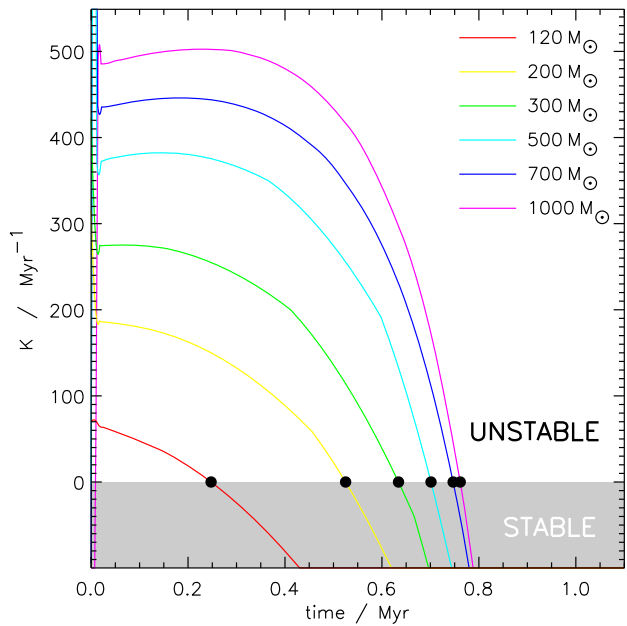}
\caption{Growth rate, $K$, as a function of the age of the star.  The
amplitude of the pulsation grows with time as
$A(t)=A_0\exp\{\frac{2\pi i}{P}t\}\exp\{Kt\}$ \cite{BHW01}.
\label{K}}
\end{figure}

Stars above $\sim60\,\Msun$ may also be unstable to the epsilon
mechanism - pulsations driven by the high temperature sensitivity of
nuclear burning \cite{Led41,SH59}.  However, historical studies of
stellar stability focused on stars of solar composition and did not
take into account the peculiarities of the first generation of stars.
We have recently studied the pulsational instability due to epsilon
mechanism in massive metal-free stars between 120 and 1000\,\Msun
\cite{BHW01}.  Their structure is uniquely different from later
stellar generations.  Since hydrogen burning through the ``pp-chain''
is not sufficient, the star contracts to a high enough temperature to
produce carbon by the triple-alpha reaction, more than 1\EE8\,K in the
stars investigated here.  A mass fraction of $\sim$1\EE{-9} is
sufficient to stop the contraction and supply energy by the CNO cycle
for hydrogen burning.  The stars, however, stay very compact and hot
in their centers throughout their hydrogen burning lifetime.  The
higher temperature causes a lower temperature sensitivity of the
nuclear energy generation.  From pulsational analysis we find that the
epsilon mechanism is weak in these compact stars and operates only for
a fraction of the hydrogen-burning life-time (Fig.~\ref{K})
\cite{BHW01}.  Radial pulsations driven by opacity or recombination
are not found.  We estimate that the resulting mass loss due to the
epsilon mechanism should be less than 5\,\% for a 500\,\Msun star,
and perhaps 10\,\% for a 1000\,\Msun star, but quite negligible for
stars of lower mass.  Stars of $\sim500\,\Msun$ or more may encounter
a red supergiant phase towards the end of central helium burning.
This could result in an additional, maybe significant, mass loss, but
its strength is not yet known.  Therefore stars of
$\lesssim500\,\Msun$ can be safely assumed to reach carbon burning
without significant mass loss.

\section{Supernovae at the Edge of the Universe}
\label{z20}

\begin{figure}[t]
\includegraphics[angle=90,width=\columnwidth]{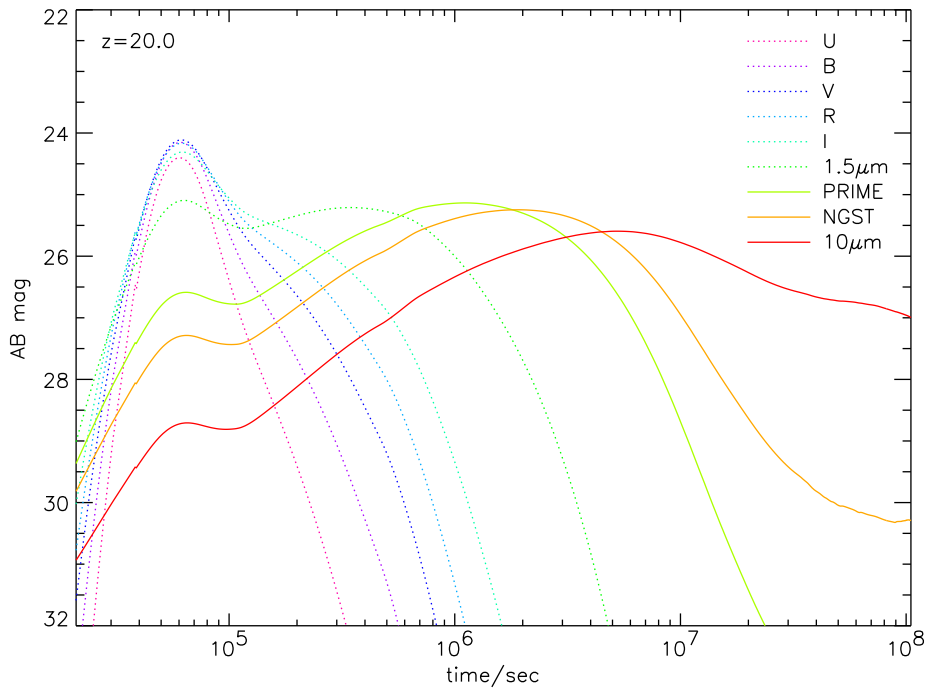}

\caption{Preliminary light curve of pair-creation supernova from a
250\,\Msun star at $z=20$ as computed by the KEPLER code \cite{WZW78}.
Time, wave lengths and magnitudes (without internal or intergalactic
extinction) are given in observer rest frame.  Wave lengths that are
beyond the IGM Ly--$\alpha$ absorption (2.55$\mu$m) are displayed as
\textsl{dotted lines}.  ``PRIME'' and ``NGST'' corresponds to 3.5 and
5.0$\mu$m.  The ``spherically symmetric'' emission has been folded to
account for the extent of the ``photosphere''.  The first ``bump'' at
$\sim\Ep3\,$s is from the shock breakout, the right ``peak'' is the
peak of the SN light curve.
\label{L}}
\end{figure}

Stars that reach carbon burning with a helium core mass of $\sim 64
\ldots 133$\,\Msun (corresponding to ZAMS masses of $\sim 140 \ldots
260$\,\Msun for stars without mass loss) are unstable to pair-creation
(see Fig.~\ref{MM3}), collapse, then explode as a supernova
(SN). These are the most powerful thermonuclear explosions in the
universe.  Pair-creation SNe release energies ranging from
3\EE{51}\,erg for a 64\,\Msun He core up to almost 100\EE{51}\,erg for
a 133\,\Msun He core \cite{HW01} -- enough energy to disrupt a small
proto-galaxy.  In Fig.~\ref{L} we show, in the observer frame, the
early light curve of an exploding 250\,\Msun star, neglecting
intergalactic and interstellar absorption.  This star has a He core of
120\,\Msun and a total explosion energy of 65\EE{51}\,erg, producing
21\,\Msun of \II{56}{Ni}.  The visible brightness is not quite as
impressive since most of the energy is kinetic, but nevertheless
should be a few times brighter than a typical Type~Ia SN.  The
bolometric luminosity of an event at $z=20$ is not much dimmer than at
a red shift of a few, where Type Ia SNe have already been observed.
The main difference, however, is the significantly larger red shift
and time dilation.  Given the bigger intrinsic time scale associated
with the large ejected mass, they also last much longer.

For a current standard cosmology ($\Ol=0.7$, $\Om=0.3$,
$\Ho=65$\,km/s/Mpc, $\Ob=0.02/h^2=0.047$) and assuming
that $\fst=\Ep{-6}$ of all baryons goes into stars of
$\Mst=250\,\Msun$, at a red shift of $z=20$, we estimate the
pair-creation supernova rate by
\[
\rpsn=4\pi\left(\frac{\dL}{1+z}\right)^{\!\!2}
\frac{c}{1+z}\rhoc\Ob(1+z)^3\frac{\fst}{\Mst}
     =\frac{4\pi\dLs c\rhoc\Ob\fst}{\Mst}
\]
where $\rhoc=3\Ho^2/8\pi G$, and the luminosity distance
$\dL=243$\,Gpc.  This gives $\sim$0.16 events per second per universe,
i.e., $\sim$3.9\EE{-6} events per second per square degree.  The first
peak of the light curve (Fig.~\ref{L}) lasts for about a month, the
second peak (``plateau''; not depicted here), probably brighter in the
far infrared, would peak for about 10\,yr (based on preliminary
calculations by Phil Pinto, priv.\ com.).  Statistically, at any time
per square degree about a dozen of these supernovae should be at the
peak of the light curve, and more than $\sim$1000 in the plateau phase
of the light curve.  Note that $\dL\propto z$ for high $z$ and thus
this rate depends critically on the red shift adopted as well as on
the baryon fraction assumed to go into these stars.

\section{Nucleosynthesis}
\label{nuc}

\begin{figure}[t]
\includegraphics[angle=90,width=\columnwidth]{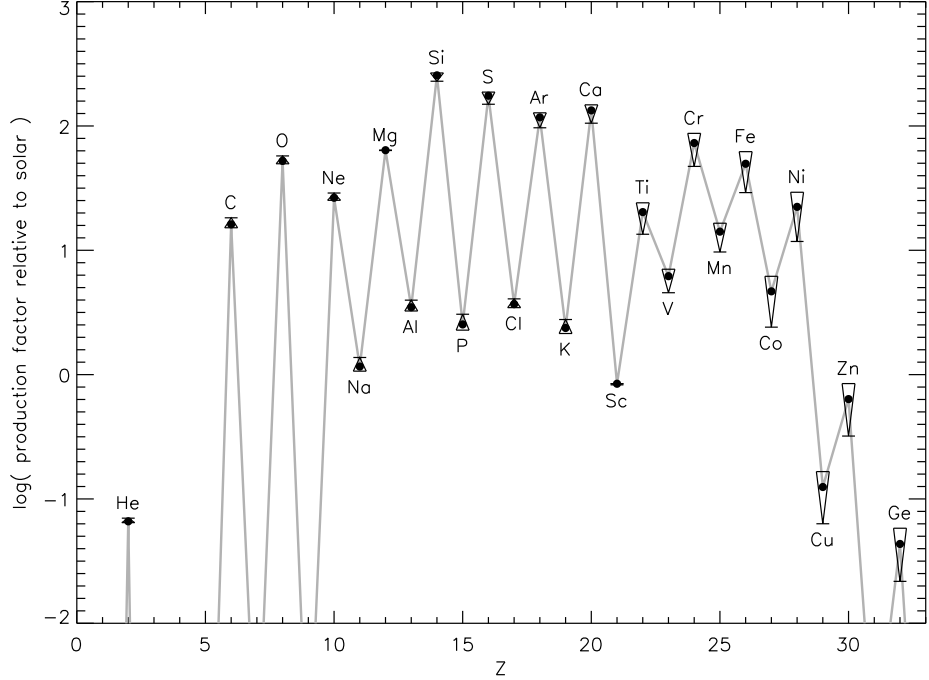}
\caption{Elemental production factors relative to solar in Pop III
pair-creation supernovae (helium core mass 65--130\,\Msun).  The thin
and thick ends of the triangle and the dots correspond to IMF slopes
of $\mathrm{d}\log{N}/\mathrm{d}\log{M}=-1.5$, $-4.5$, and $-2.5$,
respectively.\label{2snnuc}}
\end{figure}

In massive stars most of neutrons responsible for the ``weak
component'' of the $s$-process are made from initial CNO seeds during
helium burning.  Since Pop III stars are devoid of these seeds, both
the $s$-process and the production of a ``neutron excess'' for the
advanced burning stages is strongly suppressed \cite{tru}.  In massive
stars the odd-$Z$ elements are therefore underproduced with respect to
the $\alpha$ elements \cite{HW01}.  Above $\sim25$\,\Msun significant
fallback onto the remnant already occurs after the SN \cite{Fry99}
(Fig.~\ref{MM3}), and few or no heavy elements can be ejected.  From
$\sim40$ to $\sim100$\,\Msun a black hole is formed directly
\cite{Fry99} and essentially the whole star will be swallowed --- no
nucleosynthesis products are ejected.  Between $\sim100$ and
$\sim140$\,\Msun the pulses of pair instability \cite{BAC84} will
eject the outer layers of the star, possibly including parts of the CO
core, especially at the high-mass end of the regime, but nothing
heavier than magnesium leaves the star.  After the pulses these
objects probably encounter the same fate as their lighter cousins,
i.e.,
the remaining part of the star falls into a black hole.  For initial
masses of $\sim140 \ldots 260\,\Msun$, the pair instability
completely disrupts the star when explosive burning of oxygen and
silicon release enough energy to reverse the collapse into an
explosion.  However, this explosion is too rapid and the star not
dense enough at the point when the implosion turns around to lead to
significant neutronization.  A very marked odd-even pattern results
(Fig.~\ref{2snnuc}) with silicon being the biggest overproduction and
elements above germanium are essentially not produced.  In stars more
massive than $\sim260$\,\Msun, photo-disintegration of nuclei becomes
important and the collapse is not reversed.  The whole star falls into
a black hole (Fig.~\ref{MM3}).

\section{Conclusions}
\label{end}

Due to the absence of metals the first generation of star will likely
not experience significant mass loss by radiation-driven stellar winds
or opacity-driven pulsations.  Their unique structure also prevents
significant mass loss by the epsilon mechanism.  Therefore
very massive Pop III single stars reach carbon burning
with enough mass to encounter the pair instability.  Since
current theoretical studies indicate that such star may constitute a
significant, if not dominant, fraction of Pop III, we predict that
their nucleosynthetic yields may have a unique imprint on the chemical
evolution of the early universe.  Their production of odd-$Z$ elements
is by $\sim2$ orders of magnitude lower than that of even-$Z$.
Elements heavier than zinc are not produced.

Since stars of lower mass ($\lesssim 140\,\Msun$) or higher mass
($\gtrsim 260\,\Msun$) collapse into black holes without significant
heavy element creation. Pair-creation SNe are thus a ``clean''
source of nucleosynthesis in the sense that neighboring mass ranges do
not ``pollute'' the sample.  In case of a bimodal Pop III IMF, massive
stars in the range $\sim8\ldots40\,\Msun$ will also contribute to the
resulting abundance pattern, though on a slightly slower time-scale
(factor $\sim2$).  Also in them, the lack of initial CNO ``seeds''
will lead to an elemental odd-even pattern, though much less expressed
\cite{HW01}, and they possibly contribute $r$-process isotopes.  It is
even conceivable that Pop III AGB stars contribute some $s$-process
\cite{ven01}.  The interaction of the ejecta with the surrounding
matter, a possible enrichment of the intergalactic medium and mixing
of contributions from different-mass sources before the formation of
the first Pop II stars will have to be studied in more detail in the
future.

We predict that Pop III pair-creation SNe might be detectable by
future near infrared space experiments --- all the way out to the edge
of universe --- to redshifts of 20 or more.  Combined with the
challenge to find old Pop II stars that show the predicted abundance
pattern from the ashes of these explosions, this should allow deeper
insight into the happenings at the times when the first sparks of
stellar light terminated the ``dark ages''.

{\footnotesize \textbf{Acknowledgments} This research has been
supported by the NSF (AST 97-316569), the DOE ASCI Program (B347885),
the DOE SciDAC Program, and the Alexander von Humboldt-Stiftung
(FLF-1065004).}

%

\end{document}